\documentclass{article}
\usepackage{spconf,amsmath,graphicx}
\usepackage{times}
\usepackage{epsfig}
\usepackage{amssymb}
\usepackage{booktabs}
\usepackage{url}
\usepackage{hyperref}
\usepackage{color}

\title{SPECTRON: TARGET SPEAKER EXTRACTION USING CONDITIONAL TRANSFORMER WITH ADVERSARIAL REFINEMENT}
\name{Tathagata Bandyopadhyay}
\address{Visual Computing Lab, Technical University of Munich, Germany \\tathagata.bandyopadhyay@tum.de}
\begin{document}
%
\maketitle
\begin{abstract}
Recently, attention-based transformers have become a de facto standard in many deep learning applications including natural language processing, computer vision, signal processing, etc.. In this paper, we propose a transformer-based end-to-end model to extract a target speaker's speech from a monaural multi-speaker mixed audio signal. Unlike existing speaker extraction methods, we introduce two additional objectives to impose speaker embedding consistency and waveform encoder invertibility and jointly train both speaker encoder and speech separator to better capture the speaker conditional embedding. Furthermore, we leverage a multi-scale discriminator to refine the perceptual quality of the extracted speech. Our experiments show that the use of a dual path transformer in the separator backbone along with proposed training paradigm improves the CNN baseline by $3.12$ dB points. Finally, we compare our approach with recent state-of-the-arts and show that our model outperforms existing methods by $4.1$ dB points on an average without creating additional data dependency.
\end{abstract}
\section{Introduction}
\label{sec:intro}

Blind source separation \cite{belouchrani1998blind,naik2014blind} or more specifically speech separation \cite{makino2007blind,wang2018supervised}, in general refers to splitting a mixed signal to all of its constituting component audios; i.e., separating out all the speakers from a mixed speech. With the advent of deep learning, many different approaches and corresponding architectures that use convolutional \cite{luo2019conv}, recurrent \cite{luo2020dual}, or transformer \cite{chen2020dual,subakan2021attention} models have been proposed to tackle single channel speech separation. Inspired from two breakthrough papers ``TasNet" \cite{luo2018tasnet} and ``ConvTasNet" \cite{luo2019conv}, the majority of supervised speech separation approaches follow a common high-level structure including a \emph{waveform encoder} to transform the audio input to a spectrogram-like latent representation, a \emph{separator backbone} to separate the speech sources in latent space, and then a \emph{waveform decoder} to generate the wave forms for individual speakers.

General speech separation for known and small number of speakers has seen great success in recent years with deep neural networks; e.g., ``SepFormer" \cite{subakan2021attention}, ``DPTNet" \cite{chen2020dual}, ``Sandglasset" \cite{lam2021sandglasset}, ``Conv-Tasnet" \cite{luo2019conv}. However, knowing the number of speakers a priory in real-world mixtures is an impractical assumption. This is further exacerbated as there is no easy way to resolve the ambiguity between the separated channels due to lack of specific ordering amongst the speakers.

On the contrary, often many downstream applications such as a voice activated virtual assistant need to extract the speech of a pre-enrolled specific speaker from mixed speech audio inputs, where it is quite practical to assume, without loss of generality, to have the clean reference speech sample pre-recorded during speaker enrollment. In the literature this task is referred to as target speaker extraction.

In this work, we propose a target speaker extraction system which takes a monaural multi-talker speech mixture and a clean reference speech sample of the target speaker as inputs and extracts out the target speaker's speech from the mixture. A state-of-the-art speaker verification module \cite{wan2018generalized} is used to obtain the representative speaker embedding which is fed into the separator backbone as a condition alongside the spectrogram-like representation of the input speech mixture. The separator backbone conditionally estimates a mask on the transformed representation to suppress the interfering speeches. Finally, a waveform decoder is used to generate clean speech waveform\footnote{Output samples: \url{https://tatban.github.io/spec-res/}} from this masked representation. 

Like most of the existing methods for speech separation or speaker extraction \cite{luo2018tasnet, luo2019conv, luo2020dual, chen2020dual, zhang2020x}, we learn an internal representation based on a masking approach and further leverage a speaker extraction framework consisting of a speaker encoder and a separator module. However, instead of defining custom speaker encoder as in \cite{li2020atss} or custom separation module as in \cite{wang2018voicefilter}, we use a current SOTA speaker verification module \cite{wan2018generalized} as speaker encoder and adapt one of the transformer based general speech separation SOTA models known as ``DPTNet" \cite{chen2020dual} for separation module, to combine the best of both worlds.
\begin{figure*}[!ht]
    \centering
    \includegraphics[width=\textwidth,height=0.5181\columnwidth]{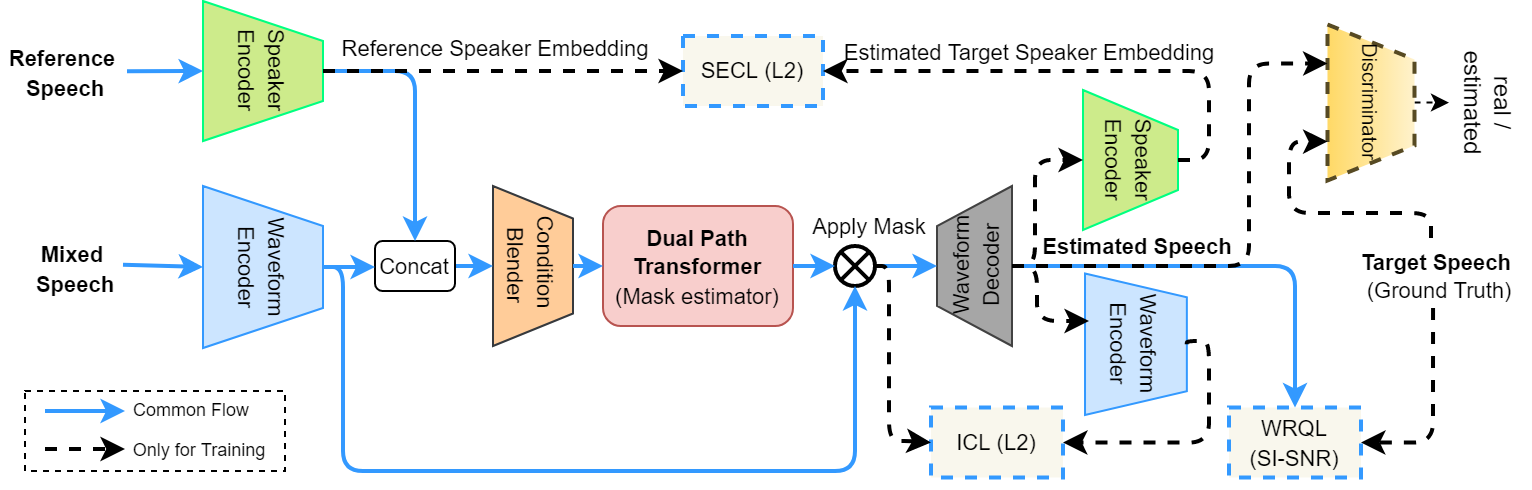}
    \caption{
    Spectron framework: same color blocks refers to shared weights; dashed boxes refers to objective functions. Speaker Embedding Consistency Loss (SECL) and waveform encoder decoder Inverse Consistency Loss (ICL) are realized as MSE Loss, where SI-SNR \cite{le2019sdr} is used for Waveform Reconstruction Quality Loss (WRQL) and ``MSD" \cite{kong2020hifi} for discriminator loss.}
    \label{fig:method}
\end{figure*}
 Like many speaker extraction systems \cite{luo2019conv, zhang2020x, chen2020dual}, we too use SI-SNR \cite{le2019sdr} loss as our main \emph{Waveform Reconstruction Quality Loss} (WRQL). However, unlike the existing approaches, we introduce \emph{Inverse Consistency Loss} (ICL) to impose invertibility of learned waveform encoder and decoder and \emph{Speaker Embedding Consistency Loss} (SECL) to impose similarity between reference speaker sample and extracted speech embeddings. Finally, we use a \emph{Multi-Scale Discriminator} (MSD) from \cite{kong2020hifi} to further improve the perceptual quality of the extracted speech.
\section{RELATED WORK}
\label{sec:related_work}

\textbf{Speech Non-Speech separation.} This is also known as speech enhancement or speech denoising \cite{li2021real, tzinis2022remixit, isik2020poconet, hao2021fullsubnet} and works with single speaker assumption. In contrast, our method works with multi-speaker assumption and extracts the clean speech of the speaker of interest.

\textbf{Multi-Speech Separation.} It assumes there are n $(n>1)$ speakers speaking simultaneously and the goal is to separate out all the speakers into n different channels at once \cite{subakan2021attention, chen2020dual, luo2020dual, luo2019conv, luo2018tasnet}. Our approach is different as we aim to conditionally extract out only the target speaker's speech.

\textbf{Target Speaker Extraction.} This is the general category of our approach and there are quite a few existing works with common high level structure consisting of a speaker encoder module and a speech separation module. Voice Filter \cite{wang2018voicefilter} uses a pre-trained speaker verification model from \cite{wan2018generalized} as a fixed speaker encoder along with a fixed time frequency representation (STFT and inverse STFT) based CNN LSTM architecture which involves complex phase estimation and long term sequential dependency modeling, making it inefficient and limited in performance. Atss-Net \cite{li2020atss} improves over it by using multi-head attention based separator jointly trained with ResNet-18 based speaker encoder. However, this approach still suffers from complex phase estimation related shortcomings due to STFT and inverse STFT. X-TasNet \cite{zhang2020x} significantly pushed further previous two SOTAs by using learnable time frequency like representation as first proposed in \cite{luo2019conv}. However, CNN based separator architecture is not efficient enough to simultaneously capture utterance level and long term dependencies due to fixed receptive field of the convolution kernels. Our approach differs from existing approaches as we utilize five things together namely: i) \emph{learned spectrogram-like representation}, ii) \emph{transformer based separator backbone}, iii) \emph{joint training of speaker encoder and speech separator}, iv) \emph{introduction of two additional objectives to impose speaker embedding consistency and waveform encoder invertibility} and v) \emph{use of multi-scale discriminator (``MSD") to refine the extracted speeches}.
\section{METHOD}
\label{sec:method}
A schematic diagram of our proposed pipeline is shown in Fig. \ref{fig:method}. We first discuss the Spectron architecture and finally elaborate on design of different objective (loss) functions involved in training.
\subsection{Model Architecture}
Spectron consists of two high level components namely \emph{speaker encoder} and \emph{speech separator}. 
\subsubsection{Speaker Encoder}
This block is responsible for computing a speaker representative embedding $e\in\mathbb{R}^{d}$ from a clean reference speech $r\in\mathbb{R}^{1\times t}$. Speaker representation learning is a crucial part of any speaker verification system. Hence, following \cite{wang2018voicefilter} and \cite{zhang2020x}, we too adopt pre-trained GE2E \cite{wan2018generalized} speaker verification model as our speaker encoder. However, unlike previous approaches, we jointly train it with the separator module. We also keep the speaker embedding dimension to $256$ not to alter the original GE2E architecture. 
\subsubsection{Speech Separator}
\label{subsubsec:Separator }
This module takes input speech mixture $i\in\mathbb{R}^{1\times t}$ along with reference speech embedding $e\in\mathbb{R}^{d}$ and conditionally estimates $\hat{s}\in\mathbb{R}^{1\times t}$, the speech of the target speaker selected by the reference speech embedding $e$. Speech Separator module of Spectron is adapted from ``DPTNet" \cite{chen2020dual} and has following sub-modules.

\textbf{Waveform Encoder} is 1D CNN with $N$ filters of kernel size $k$ and stride $st$. It is a learnable mapping from $x\in\mathbb{R}^{1\times t}$ to  $X\in\mathbb{R}^{N\times T}$ where $x$ is the input waveform and $X$ is a spectrogram-like internal representation. Following ``DPTNet" \cite{chen2020dual}, we set $N=64$, $k=16$ and $st=8$ and pass the output of this mapping through ReLU activation. 

\textbf{Condition Blender} is a 1D CNN with both kernel size and stride equal to $1$ and it takes the Waveform Encoder output concatenated with the speaker embedding to reduce its dimension to match it with the required input dimension of the Separator core. This helps us to keep the separator core architecture unaltered from ``DPTNet" \cite{chen2020dual}. Number of filters used in it is equal to the input channel dimension of the separator core which is $64$ in our case.

\textbf{Separator Core} is the same dual path transformer architecture from ``DPTNet" \cite{chen2020dual} paper with only difference in number of attention heads. We use $8$ attention heads instead of originally proposed $4$, keeping other configuration unaltered. This block estimates a target speaker specific conditional mask of same shape of waveform encoder output. This mask is multiplied element-wise with the waveform encoder output to separate out the target speaker in this transformed spectrogram-like space.

\textbf{Waveform Decoder} is intuitively an inverse mapping of the waveform encoder to produce the target speech waveform from the masked internal representation. However, instead of strictly enforcing this inverse property, we keep it as learnable just like ``DPTNet" \cite{chen2020dual} but introduce an additional loss function to softly impose invertibility.
\subsection{Design of Training Objectives}
We use four objective (loss) functions as shown with dashed boxes in Fig. \ref{fig:method}. Descriptions of those are as follows:
\subsubsection{Waveform Reconstruction Quality}
As the main goal of speaker extraction is to extract out the clean speech of the target speaker from the input mixture, we need to improve the waveform reconstruction quality (WRQ) by maximizing signal-to-noise ratio (SNR) of the estimated speech. To avoid the scale dependency we use SI-SNR (scale invariant SNR also known as SI-SDR) as proposed by \cite{le2019sdr} and use negative value of it to formulate as minimization objective. Let us consider $s,\hat{s}\in\mathbb{R}^{1\times t}$ represent ground truth and estimated speech signals respectively and both of them are normalized to zero mean. Then, SI-SNR and WRQL can be formulated as: 
\begin{equation}
s_{\text {target }}=\frac{\langle\hat{s}, s\rangle s}{\|s\|^{2}}
\end{equation}
\begin{equation}
e_{\text {noise }}=\hat{s}-s_{\text {target }}
\end{equation}
\begin{equation}
SI\mbox{-}SNR:=10 \log _{10} \frac{\left\|s_{\text {target }}\right\|^{2}}{\left\|e_{\text {noise }}\right\|^{2}}
\end{equation}
\begin{equation}
    WRQL:=-SI\mbox{-}SNR
\end{equation}
\subsubsection{Speaker Embedding Consistency}
As per our problem formulation, the input reference speech and the estimated target speech, even though content wise different, are spoken by the same speaker. Therefore, both of them should have similar voice textures which means they should produce similar embedding vectors when passed through speaker encoder. With this intuition we formulate the speaker embedding consistency loss (SECL) as follows:
\begin{equation}
    SECL:=\|SE_\theta(r)-SE_\theta(\hat{s})\|^{2}
\end{equation}
where $r,\hat{s}\in\mathbb{R}^{1\times t}$ represent reference and estimated speeches respectively and $SE_\theta:\mathbb{R}^{1\times t}\xrightarrow{}\mathbb{R}^{d}$ is the Speaker Encoder parameterized by $\theta$, which produces fixed length embedding from variable length speech samples.
\subsubsection{Inverse Consistency}
Waveform encoder (WE) and waveform decoder (WD) should be inverse operation of each other with an intuitional analogy of STFT and inverse STFT. We introduce inverse consistency loss (ICL) to implicitly impose this constraint. Let $m\in\mathbb{R}^{N\times T}$ denotes the masked representation of the separated speech in spectrogram-like transformed space. Then ICL can be formulated as:
\begin{equation}
    ICL:=\|m-WE_\gamma(WD_\delta(m))\|^{2}
\end{equation}
where $WE_\gamma:\mathbb{R}^{1\times t}\xrightarrow{}\mathbb{R}^{N\times T}$ and $WD_\delta:\mathbb{R}^{N\times T}\xrightarrow{}\mathbb{R}^{1\times t}$ are waveform encoder parameterized by $\gamma$ and waveform decoder parameterized by $\delta$ respectively.
\subsubsection{Adversarial Refinement}
Finally, we use a multi-scale discriminator (MSD) \cite{kong2020hifi, kumar2019melgan} in an adversarial setting with a goal to make the estimated target speech indistinguishable from the ground truths. We train MSD to classify the ground truth samples to class $1$ and estimated samples to class $0$, where as the generator i.e the Speech Separator (See \ref{subsubsec:Separator }) is trained to fool the discriminator. The adversarial losses can be formulated as:
\begin{equation}
    \mathcal{L}_{\text{d}}(D;G) := \mathbb{E}_{(s,i,e)}\left[(D(s)-1)^2+(D(G(i,e)))^2\right]
\end{equation}
\begin{equation}
    \mathcal{L}_{\text{g}}(G;D) := \mathbb{E}_{(i,e)}\left[(D(G(i,e))-1)^2\right]
\end{equation}
where $G:\mathbb{R}^{1\times t}\times \mathbb{R}^{d}\xrightarrow{}\mathbb{R}^{1\times t}$ denotes generator i.e speech separator, $D:\mathbb{R}^{1\times t}\xrightarrow{}\{0,1\}$ denotes discriminator and $s,i,e$ carry previously defined meanings. $\mathcal{L}_{\text{g}}(G;D)$ is added with the other three losses and $\mathcal{L}_{\text{d}}(D;G)$ is optimized with a second optimizer (AdamW) with same learning rate.
\section{RESULTS}
\label{sec:results}
\subsection{Dataset}
We base our experiments on different mixture subsets generated from LibriSpeech\footnote{\url{http://www.openslr.org/12/}} data \cite{panayotov2015librispeech}. In particular, for ablation we use LibriMix \cite{cosentino2020librimix} script\footnote{\url{https://github.com/JorisCos/LibriMix}} on \emph{``train-clean-100"}, \emph{``dev-clean"} and \emph{``test-clean"} for creating training, validation and test mixtures respectively. We collectively refer this data as ``LibriMix Data". On the other hand, for SOTA comparison we use the same mixture subsets released by google\footnote{\url{https://tinyurl.com/2vc795dw}} and already used in \cite{wang2018voicefilter, li2020atss, zhang2020x}. We refer this dataset as ``VoiceFilter Data".
\subsection{Experimental Setup}
For all the experiments we have used $batchsize=4$, $learning rate=1e^{-4}$, $weight decay=1e^{-7}$ and train with Adam (and AdamW for discriminator) optimizer(s) for $201$ epochs and use the best validation weights to compute the performance measures on the test set. Our speech separator module operates at $8$ KHz sampling frequency, where as speaker encoder uses $16$ KHz. Therefore, re-sampling is taken care on the fly with torchaudio defaults.

It is also noteworthy, that for ``LibriMix Data" we split the clean speeches in non-overlapping $3$ second and $2$ second segments to use them as ground truth and reference speech respectively as we don't have separate reference speech. However, for ``VoiceFilter Data", as we have mix, ground truth and reference speeches we use $3$ second segment for each of them.
\subsection{Ablation Study}
\begin{table}[]
\centering
\resizebox{\columnwidth}{!}{%
\begin{tabular}{@{}lll@{}}
\toprule
Model Variant                     & SDRi (dB)      & SI-SNRi (dB)   \\ \midrule
Baseline                          & 11.13          & 10.42          \\
Baseline+ICL                      & 10.92          & 10.07          \\
Baseline+ICL+SECL                 & 10.95          & 10.15          \\
Baseline+ICL+SECL+JointTraining   & 12.41          & 11.72          \\
DPTNet+ICL+SECL+JointTraining   & 13.94          & 13.23          \\ \midrule
Spectron (with ``DPTNet" and ``MSD") & \textbf{14.25} & \textbf{13.44} \\ \bottomrule
\end{tabular}%
}
\caption{Ablation study of Spectron on 2 speaker mixtures.}
\label{tab:ablation}
\end{table}
We show ablation study of Spectron in Table \ref{tab:ablation}. We start with a naive CNN baseline with fixed pre-trained Speaker Encoder\footnote{pre-trained encoder from: \url{https://tinyurl.com/4r4a63np}} and ``ConvTasnet" Separator module trained with only Waveform Reconstruction Quality Loss (WRQL). Then, we introduce Inverse Consistency Loss (ICL) followed by Speaker Embedding Consistency Loss (SECL). Training the speaker encoder jointly instead of keeping it fixed improves the baseline performance by roughly 1.3 dB points. This is probably because joint training allows to capture better relationship between the speaker embedding computation and its use in conditional speech separation. Finally, introduction of ``DPTNet" separator backbone and multi-scale discriminator (MSD) loss, along with previous losses and joint training, improves the baseline by 3.12 dB points.
\subsection{Comparison with Existing Works}
\begin{table}[]
\centering
\resizebox{\columnwidth}{!}{%
\begin{tabular}{@{}lll@{}}
\toprule
Model               & SDRi (dB) & SI-SNRi (dB) \\ \midrule
VoiceFilter \cite{wang2018voicefilter}         & 7.8       & -            \\
AtssNet \cite{li2020atss}             & 9.3       & -            \\
X-Tasnet \cite{zhang2020x}           & 13.8      & 12.7         \\ \midrule
Spectron without MSD (ours)     & 13.9      & 12.8         \\
Spectron (ours)     & \textbf{14.4}      & \textbf{13.3}         \\ \bottomrule
\end{tabular}%
}
\caption{Spectron performance vs recent state-of-the-arts.}
\label{tab:sota}
\end{table}
We compare quantitative performance of Spectron with existing state-of-the-art target speaker extraction frameworks in Table \ref{tab:sota}. To make a fair comparison across models, here we use same train and same test data sets as used in \cite{wang2018voicefilter, li2020atss, zhang2020x}. As we see, Spectron undoubtedly performs better than ``VoiceFilter" \cite{wang2018voicefilter}, ``AtssNet" \cite{li2020atss} and naive ``X-Tasnet" \cite{zhang2020x}. However, a variant of ``X-Tasnet" which uses \emph{Loss on Distortion} (LoD), performs slightly ($+0.3$ dB points) better than Spectron, but at the cost of significant amount of additional data and more complicated training procedure as LoD needs ground truth speeches of all the speakers in the mixture.

\section{Conclusion}
We have presented Spectron, which uses dual path transformer conditioned on speaker embedding produced by a speaker encoder to extract the speaker of interest. Attention mechanism in the transformer, introduction of two additional  objective functions followed by adversarial refinement and joint training of speaker encoder and speech separator - these four ideas all together improves the baseline performance and thus push the current SOTA further in target speaker extraction. Spectron can be directly used in different down stream speech based applications like automatic speech recognition, conditional speaker diarization, voice command activated personal assistants and so on. Additionally, it enables interactive audio manipulation, where the clean portions of an audio can be used as reference to de-noise the cluttered  portions of the same audio. In future, Spectron can be explored as a general framework for any kind of target audio extraction with suitable reference audio encoder.

\vfill\pagebreak
\section{Acknowledgement}
This project was done at the Visual Computing and Artificial Intelligence lab of Technical University of Munich under the kind supervision of Prof. Dr. Matthias Niessner. Author cordially thanks all lab members for useful discussion and timely co-operation. 
{\small
\bibliographystyle{IEEEbib}
\bibliography{ref}
}

\end{document}